\documentclass[twocolumn,aps,prb,superscriptaddress,amsmath,amssymb,showpacs,preprintnumbers,floatfix]{revtex4-1}

\usepackage{graphicx}
\usepackage{natbib,color,bm}
\usepackage{multirow}

\def\kk{\ensuremath{\mathbf q}}
\def\tb{\ensuremath{\mathbf t}}
\def\Qb{\ensuremath{\mathbf Q}}

\def\ab{\ensuremath{\mathbf{a}^\ast}}
\def\bb{\ensuremath{\mathbf{b}^\ast}}
\def\cb{\ensuremath{\mathbf{c}^\ast}}
\def\qicm{\ensuremath {\mathbf q}_{\rm icm}}
\def\qcm{\ensuremath {\mathbf q}_{\rm cm}}

\begin{document}



\title{Electric field control of chiral magnetic domains in the high-temperature multiferroic CuO}

\author{P.~Babkevich}
\email[]{peter.babkevich@physics.ox.ac.uk}
\affiliation{Department of Physics, Oxford University, Oxford, OX1 3PU, United Kingdom}
\author{A.~Poole}
\affiliation{Laboratory for Neutron Scattering, Paul Scherrer Institut, CH-5232 Villigen PSI, Switzerland}
\author{R.~D.~Johnson}
\affiliation{Department of Physics, Oxford University, Oxford, OX1 3PU, United Kingdom}
\affiliation{ISIS facility, Rutherford Appleton Laboratory-STFC, Chilton, Didcot, OX11 0QX, United Kingdom}
\author{B.~Roessli}
\affiliation{Laboratory for Neutron Scattering, Paul Scherrer Institut, CH-5232 Villigen PSI, Switzerland}
\author{D.~Prabhakaran}
\affiliation{Department of Physics, Oxford University, Oxford, OX1 3PU, United Kingdom}
\author{A.~T.~Boothroyd}
\affiliation{Department of Physics, Oxford University, Oxford, OX1 3PU, United Kingdom}

\date{\today}

\begin{abstract}
We have studied the high temperature multiferroic cupric oxide using polarized neutron diffraction as a function of temperature and applied electric field. We find that the chiral domain population can be varied using an external electric field applied along the $b$ axis. Using representation analysis we derive the incommensurate magnetic structure in the multiferroic phase. The origin of the magnetoelectric coupling is consistent with models based on the inverse Dzyaloshinskii--Moriya interaction, but is different from the simple cycloidal mechanism.
\end{abstract}

\pacs{75.85.+t, 75.25.-j, 77.80.-e, 61.05.F-}


\maketitle



\section{Introduction}
Materials in which ferroelectric and magnetic order parameters couple are of great interest for application in spintronic devices. Such materials are termed \emph{multiferroic} and have been actively studied in recent years.\cite{wang-advphys-2009, cheong-nature-2007, khomskii-physics-2009} Of particular interest are the type-II multiferroics in which the magnetic order is directly and strongly coupled to the electric polarization. The strong coupling is often associated with non-collinear magnetic structures which originate from competition and frustration in the magnetic interactions. Most of the strongly-coupled multiferroics, therefore, exhibit their unusual properties at low temperatures ($< 40$\,K).  An exception is cupric oxide, which develops a multiferroic phase below 230\,K.\cite{kimura-nature-2008}

The magnetic phase diagram of CuO in zero magnetic field contains two phases, AF1 and AF2.\cite{brockhouse-pr-1954, forsyth-jphysc-1988, ain-jpcm-1992, brown-jpcm-1991, yang-prb-1989} The low temperature paraelectric phase (AF1) is a simple commensurate antiferromagnet which is stable up to $T_{\rm N1} = 213.7$\,K. At $T_{\rm N1}$ there is a first-order transition to the multiferroic phase (AF2), which has a complex, non-collinear incommensurate magnetic structure and a ferroelectric polarization along the crystallographic $b$ axis. Both magnetic and ferroelectric orders vanish above $T_{\rm N2} = 230$\,K.

The mechanism that causes the simultaneous emergence of ferroelectricity and magnetic ordering in CuO is still under discussion and a number of theories have recently been proposed.  A phenomenological Landau theory approach was employed by Toledano {\it et al.}~(Ref.~\onlinecite{toledano-prl-2011}) to investigate the two ordered phases. Using this theory they were able to explain the sequence of phases, with the incommensurate multiferroic phase at a higher temperature than the simple antiferromagnetic phase, and predicted that both phase transitions are first order.  Giovannetti {\it et al.}~(Ref.~\onlinecite{giovannetti-prl-2011}) and Jin {\it et al.}~(Ref.~\onlinecite{jin-arxiv:1007.2274v3}) performed density functional calculations which suggest that the ferroelectric polarization is induced via the Dzyaloshinskii--Moriya (DM) interaction\cite{dzyaloshinskii-1964, moriya-pr-1960} between neighboring chains of non-collinear spins. They further argue that the small incommensurability in the AF2 phase does not play an important role in the multiferroic mechanism.

Overall, it is widely accepted that the competing exchange interactions together with geometrical frustration lead to a non-collinear magnetic ordering which breaks inversion symmetry and allows polar lattice distortions. What is different about CuO is that a strong Cu--O--Cu superexchange interaction along the $[1,0,\bar{1}]$ direction allows it to remain ordered at high temperatures.\cite{shimizu-prb-2003, yang-prb-1989, ain-physicaC-1989, boothroyd-physicaB-1997}

If the magnetoelectric coupling is strong enough, it should be possible to vary the magnetic domain population with an external electric field. The domain population can be monitored by polarized neutron or photon diffraction. This type of experiment has been performed previously on several different multiferroics, and as well as demonstrating the manipulation of magnetic domains by electric fields these experiments have provided useful information about the magnetoelectric coupling mechanisms.\cite{cabrera-prl-2009, poole-jpcs-2009, seki-prl-2008, yamasaki-prl-2007, fabrizi-prl-2009, radaelli-prl-2008}

Here we present the results of a polarized neutron diffraction study of the magnetic response of CuO in an applied electric field. We demonstrate that the magnetic domains in CuO can be switched by an electric field applied along the ferroelectric polarization axis. The field required to reverse the domain population is found to be closely related to the temperature. However, it does not appear to be possible to transform the crystal into a single magnetic domain even at large applied fields.

\begin{figure}
\centering
\includegraphics[width=0.6\columnwidth,clip=]
{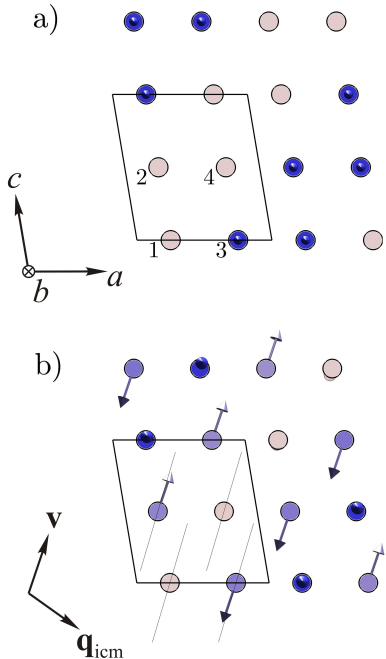}
\caption{(color online). The observed magnetic structures of CuO in projection onto the $a$--$c$ plane. The four Cu sites in the conventional structural unit cell (indicated by the parallelogram) are labelled 1--4. The colored circles represent the component of the moment along $\bf b$: light red is along $+\bf b$, dark blue is along $-\bf b$, and blue circles with arrows are moments which lie in the $a$--$c$ plane. Panel (a) shows the commensurate magnetic structure in the AF1 phase whose moments are all parallel or antiparallel to $\bf b$.  Panel (b) shows the spin arrangement in the incommensurate AF2 phase. The moments lie in the planes indicated by the dotted lines, which contain the $b$ axis and the direction $\bf v$.  The AF2 structure has two domains of opposite handedness. Only the left-handed domain is shown here.
\label{fig:crystal}}
\end{figure}


\section{Experimental procedure}

A single crystal sample of CuO was grown from a melt of high purity cupric oxide (99.995\%) by the optical floating-zone method.\cite{prabhak-jcg-2003} The crystal, which was single phase and untwinned, was cut with parallel flat faces perpendicular to the $b$ axis giving a plate-like sample of surface area of 21\,mm$^2$, thickness 0.9\,mm and mass 0.175\,g. An 8--10\,nm layer of chromium and 40--50\,nm of gold was evaporated onto the flat faces. Electrical contacts were attached using silver paste and the sample was sealed in an aluminium vacuum can to avoid electric field breakdown.

Polarized neutron diffraction measurements were performed on the TASP/MuPAD instrument at SINQ, Paul Scherrer Institut.\cite{semadeni-physicab-2001, fischer-physicab-1997} The sample was aligned with the $a$ and $c$ axes in the horizontal scattering plane such that the applied electric field was along the $-b$ direction. The scattering vector is defined as $\mathbf{Q} = \mathbf{k}_{\rm i} - \mathbf{k}_{\rm f}$ where $\mathbf{k}_{\rm i}$ and $\mathbf{k}_{\rm f}$ are the incident and scattered neutron wavevectors, and $k_{\rm i} = k_{\rm f} = 1.97$\,\AA$^{-1}$. The flipping ratio, as measured on the $(002)$ and $(200)$ structural Bragg peaks, was found to be 17. The experimental data presented here have not been corrected for the non-ideal neutron polarization, but corrections have been included in all calculations from magnetic structure models. The background signal in the different polarization channels was measured at positions away from the Bragg peaks and found to be negligible.

Pyroelectric current measurements to obtain the electric polarization of CuO were made with a customized PPMS (Quantum Design) insert. The crystal was cut from the same rod as the neutron diffraction sample. The surface area of this sample was 10\,mm$^2$ and its thickness was 1.7\,mm. The sample was cooled from 250\,K in the paraelectric phase to 220\,K in the multiferroic phase with a potential difference of 180\,V applied parallel to $b$. Upon reaching 220\,K, the electric field was turned off and the surface charge on the Cr/Au coating was allowed to discharge. The pyroelectric current was then recorded either upon warming to above $T_{\rm N2}$ or cooling down below $T_{\rm N1}$, using the same field-cool procedure before each measurement. The heating/cooling rate was 1\,K\,min$^{-1}$. To eliminate systematic errors, an applied electric field parallel and antiparallel to $b$ was employed.

Zero-field-cooled magnetization measurements were performed with a SQUID-based magnetometer (Quantum Design). The measuring field was 1000\,Oe. A slow heating rate of 0.01\,K\,min$^{-1}$ was used close to the sharp $T_{\rm N1}$ transition, and a rate of 0.1\,K\,min$^{-1}$ was used above and below it.


\section{Magnetic representation analysis}

The antiferromagnetic structure of the AF1 phase which exists below $T_{\rm N1} \approx 213$\,K is shown in Fig.~\ref{fig:crystal}(a). The magnetic propagation wavevector is $\qcm = (0.5,0,-0.5)$, and the spins lie parallel to the $b$ axis.  The multiferroic phase (AF2) exists within the narrow temperature range between $T_{\rm N1}$ and $T_{\rm N2}$. Neutron diffraction experiments have shown that the magnetic structure in this phase has an incommensurate propagation wavevector $\qicm = (0.506,0,-0.483)$. The Cu$^{2+}$ moments rotate with an approximately circular envelope in the plane containing the vectors $\bf b$ and ${\bf v} = 0.506\ab + 1.517\cb$, where \ab\ and \cb\ are reciprocal lattice basis vectors.\cite{ain-jpcm-1992,brown-jpcm-1991}  Spontaneous ferroelectric polarization is found to develop in the AF2 phase along the $b$ axis.\cite{kimura-nature-2008} No ferroelectricity is observed for $T<T_{\rm N1}$ or $T>T_{\rm N2}$.

The crystal structure of CuO can be described by the monoclinic $C2/c$ space group, with $\beta = 99.55^\circ$ and lattice parameters of $a = 4.69$\,\AA, $b = 3.42$\,\AA\ and $c = 5.13$\,\AA\ at 300\,K.\cite{yamada-prb-2004} For both magnetic phases, the magnetic propagation vector \kk\ is left unchanged by the identity $g_1 = \{E|0,0,0\}$ and glide plane $g_2 = \{\sigma_y|0,0,0.5\}$ symmetry operations of the paramagnetic space group.  The little group $G_{\bf q}$ is therefore $\{g_1, g_2\}$. The irreducible representations of $G_{\bf q}$ are $\Gamma_1$ and $\Gamma_2$, whose characters are given in Table~\ref{table:irreps}. The primitive cell of CuO contains two Cu atoms, but the $C$-centering $\{E|0.5,0.5,0\}$ then generates two more Cu sites such that there are four Cu atoms in the conventional unit cell.

\begin{table}
\begin{tabular}{c c c}
\hline
$G_{\kk}$ & $\{E|0,0,0\}$ & $\{\sigma_y|0,0,0.5\}$\\
\hline
$\Gamma_1$ & 1 & $\hphantom{-}\varphi$ \\
$\Gamma_2$ & 1 & $-\varphi$ \\
\hline
\end{tabular}
\caption{Character table of the irreducible representations of $G_{\kk}$ with the phase factor given as $\varphi = -\exp({{\rm i}\pi q_z})$.
\label{table:irreps}}
\end{table}

\begin{table}
\centering
\begin{tabular}{cccc}
\hline
& $j$ & $M_b$ & $M_{v}$\\
\hline

\multirow{2}{*}{$\Gamma_1$}
& 1 &
$\cos (\kk\cdot\tb + \theta_b)$ &
$\cos (\kk\cdot\tb + \theta_{v})$ \\

& 2 &
$\cos (\kk\cdot\tb + q_c\pi + \theta_b)$ &
$-\cos (\kk\cdot\tb + q_c\pi + \theta_{v})$ \\
\\[-2ex]

\multirow{2}{*}{$\Gamma_2$}
& 1 &
$\cos (\kk\cdot\tb + \theta_b)$ &
$\cos (\kk\cdot\tb + \theta_{v})$ \\

& 2 &
$-\cos (\kk\cdot\tb + q_c\pi + \theta_b)$ &
$\cos (\kk\cdot\tb + q_c\pi + \theta_{v})$ \\
\hline
\end{tabular}

\caption{The symmetry-allowed magnetic structures of CuO. The AF2 structure has ordering vector ${\bf q} = (0.506,0,-0.483)$ and can be resolved into components parallel to $\bf b$ and ${\bf v} = 0.506\ab + 1.517\cb$. The AF1 structure has ${\bf q} = (0.5,0,-0.5)$ and only has components along the $b$ axis. The Cu site positions $j$ are: (1) $\frac{1}{4}$,$\frac{1}{4}$,0; (2) $\frac{1}{4}$,$\frac{3}{4}$,$\frac{1}{2}$; (3) $\frac{3}{4}$,$\frac{3}{4}$,0; (4) $\frac{3}{4}$,$\frac{1}{4}$,$\frac{1}{2}$, as indicated in Fig.~\ref{fig:crystal}. The lattice translation vectors are denoted by $\bf t$, and the moments of atoms on sites 3 and 4 are found using the $C$-centering translation $\tb_C = (1/2,1/2,0)$, resulting in an additional phase of $\kk\cdot\tb_C$. For the AF1 structure, $\theta_b = \pi/4$, and for the AF2 structure, $\theta_b = 0$ and $\theta_{v} = \pi/2$.
\label{table:mag_str}}

\end{table}

The components of the two Cu spins in the primitive cell form a 6-dimensional representation $\Gamma_{\rm mag}$ of $G_{\kk}$ which reduces to $\Gamma_{\rm mag} = 3\Gamma_1 + 3\Gamma_2$. Symmetry analysis\cite{bertaut-jpc-1971} fixes the relative phases between Cu sites and the experimentally-determined spin directions further constrain the magnetic structures.  The magnetic components of the structures consistent with this analysis are given in Table~\ref{table:mag_str}.


The commensurate AF1 phase can be described using just $M_{b}$, the basis vector along $\bf b$. The spin structures corresponding to the pure $\Gamma_1$ and $\Gamma_2$ irreducible representations, $M^{(1)}_{b}$ and $M^{(2)}_{b}$, are given in Table~\ref{table:mag_str}. The choice of phase factor $\theta_b = \pi/4$ fixes the amplitude of the magnetic moment on each Cu site to be equal.\cite{ain-jpcm-1992} Single-crystal neutron diffraction measurements of magnetic Bragg peak intensities below $T_{\rm N1}$ have established that the structure for the AF1 phase has the $\Gamma_1$ symmetry.


The incommensurate AF2 phase has magnetic components along both the $\bf b$ and $\bf v$ directions, $M_{b}$ and $M_{v}$, respectively. To correctly describe the AF2 spin structure we set $\theta_b = 0$ and $\theta_{v} = \pi/2$ so that the spins rotate with a circular envelope, as observed experimentally. Assuming the $M_{b}$ and $M_{v}$ components can each be described by a single irreducible representation, $\Gamma_1$ or $\Gamma_2$, there are four possible magnetic structures which we denote by  $M^{(n)}_{b}M^{(m)}_{v}$. These structures are chiral, so each has two domains of opposite handedness related by spatial inversion. When the $b$ and $v$ magnetic components belong to the same irreducible representation, i.e.\ $M^{(1)}_{b}M^{(1)}_{v}$ or $M^{(2)}_{b}M^{(2)}_{v}$, this corresponds to a spin arrangement in which the spins on sites 1 and 3 rotate in the opposite sense along the $a$ axis to those in sites 2 and 4. This results in zero net electric polarization in the unit cell assuming the magnetoelectric coupling depends on the spin current ${\bf S}_1 \times {\bf S}_2$. The two magnetic structures with different symmetry for the $b$ and $v$ magnetic components ($M^{(1)}_{b}M^{(2)}_{v}$ or $M^{(2)}_{b}M^{(1)}_{v}$) do have a net electric polarization in the spin current model.

Of the four assumed AF2 magnetic structures for CuO, only $M^{(1)}_{b}M^{(2)}_{v}$ is consistent with neutron diffraction experiments.  The $M^{(2)}_{b}M^{(1)}_{v}$ structure can be ruled out because it predicts a very small structure factor (identically zero for an isotropic Cu form factor) for the $\Qb = \qicm$ fundamental diffraction peak, while the $M^{(1)}_{b}M^{(1)}_{v}$ and $M^{(2)}_{b}M^{(2)}_{v}$ structures give very poor agreement with neutron polarimetry measurements.\cite{footnote} The $M^{(1)}_{b}M^{(2)}_{v}$ structure is shown in Fig.~\ref{fig:crystal}(b), and has been used in all subsequent analysis presented here.

By enumerating the possible symmetry-allowed magnetic structures for the AF1 and AF2 phases of CuO and comparing these with experiment, we conclude that the structures shown in Fig.~\ref{fig:crystal} are the correct ones. We can therefore use these to quantify changes in domain population induced by an applied electric field. In the next section we describe briefly the methodology of our experiment.

\section{Spherical neutron polarimetry}
In general, magnetic scattering of neutrons involves a change in the polarization state of the neutrons. Spherical neutron polarimetry is a technique which measures the polarization of the scattered neutron beam for an arbitrary incident beam polarization. The results are expressed in terms of a polarization matrix,
\begin{equation}
\textsf{P}_{\alpha\beta} = \frac{I_{\alpha\beta}-I_{\alpha\overline{\beta}}}{I_{\alpha\beta}+I_{\alpha\overline{\beta}}},
\end{equation}
where $\alpha$ and $\beta$ indicate the polarization of the incident and final beam, respectively. $I_{\alpha\beta}$ is the magnetic Bragg peak intensity, given by
\begin{equation}
I_{\alpha\beta} = |\langle \alpha| \mathbf{M}_{\perp}({\bf Q})\cdot \bm{\sigma}| \beta\rangle |^2,
\end{equation}
where $\bm{\sigma}$ is the Pauli operator for the neutron spin and $\mathbf{M}_{\perp}({\bf Q})$ is the component of $\mathbf{M}(\Qb)$ perpendicular to \Qb, where $\mathbf{M}(\Qb)$ is the Fourier transform of the magnetization.

For CuO, $\mathbf{M}(\Qb)$ may be written,
\begin{equation}
\mathbf{M}(\Qb) = \sum_{\bf t}\sum_{j}f_j({\bf Q}){\bf M}_j({\bf t}){\rm e}^{{\rm i}{\bf Q}\cdot({\bf t}+{\bf r}_j)},\label{eq:M(Q)}
\end{equation}
where ${\bf r}_j$ is the displacement of the $j$th Cu site ($j=1$--4) from the origin of the conventional unit cell, $f_j({\bf Q})$ is the Cu magnetic form factor, and $\bf t$ is a lattice translation vector.  The components of ${\bf M}_j({\bf t})$ are given in Table~\ref{table:mag_str}.

In an ideal measurement with perfect neutron polarization the polarization matrix for CuO is given by
\begin{equation}
\textsf{P} =
\begin{pmatrix}
-1 & 0 & 0\\
 C & -D & 0\\
 C & 0 & D
\end{pmatrix},
\end{equation}
where,
\begin{equation}
C = \frac{2\Im\{M_z^{\ast}M_y\}}{|M_{\perp}|^2},\quad
D = \frac{|M_z|^2 - |M_y|^2}{|M_{\perp}|^2}.
\label{eq:polmat}
\end{equation}
Here we have used the Blume coordinate system,\cite{blume-pr-1963, izyumov-sovphys-1962} with the $x$ axis parallel to $\bf Q$, the $z$ axis perpendicular to the scattering plane, and the $y$ axis chosen to make up a right-handed set of axes. In this axis system ${\bf M}_{\perp} = (0,M_y,M_z)$. The existence of a non-zero `chiral' term $C$ in the polarization matrix implies a non-collinear magnetic structure that is, by definition, constructed from orthogonal, out-of-phase components, such as a helix or cycloid. Therefore, $C = 0$ in the collinear AF1 structure of CuO, and in the AF2 phase, left- and right-handed chiral domains will give $C<0$ and $C>0$, respectively. Equivalently, if we measure the $\textsf{P}_{yx}$ and $\textsf{P}_{zx}$ elements on a magnetic reflection we can determine the relative proportions of the two chiral domains.

The parameters $C$ and $D$ can be calculated from Eqs.~\ref{eq:polmat} for a general helicoidal magnetic structure.  For the particular case when the spins rotate with a circular envelope in a plane perpendicular to $\bf q$, the values for the reflection ${\bf Q} = {\bf q}$ are $C=\pm 1$ and $D=0$. In the AF2 phase of CuO, the propagation vector $\qicm$\ is not perpendicular to the plane of rotation of the spins, but rather lies at an angle of 107$^\circ$ to it --- Fig.~\ref{fig:crystal}(b). For ${\bf Q} = {\bf q}$ this results in a small but non-zero value of $D$, and a magnitude of $C$ which is slightly less than 1 for a single chiral domain.

The polarization matrix measured on a nuclear reflection contains only the diagonal elements $\textsf{P}_{\alpha\beta} = \delta_{\alpha\beta}$.  Measurements on nuclear (or equivalently magnetic) reflections are useful in estimating the amount of depolarization of the incident neutron beam due to stray fields. This is quantified by a flipping ratio, defined for a nuclear Bragg peak as the ratio of the scattering intensity measured in the non-spin-flip to spin-flip channel, $R = I_{\alpha\alpha}/I_{\alpha\bar{\alpha}}$.

\section{Experimental results}

\begin{figure}
\centering
\includegraphics[bb = 6 21 370 560,width=0.8\columnwidth]
{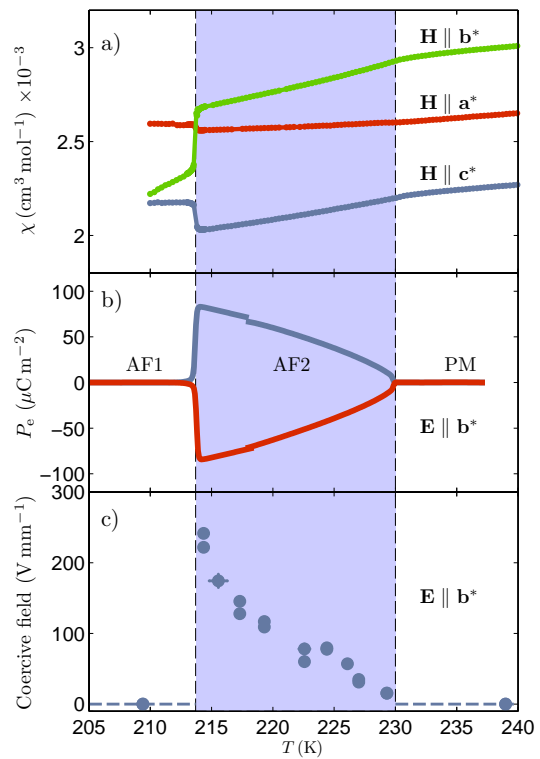}
\caption{(color online). Magnetic and dielectric properties of CuO. The upper panel (a) The zero-field-cooled magnetic susceptibility measured with a field of $H= 1000$\,Oe applied parallel to the $\{\ab,\bb,\cb\}$ reciprocal lattice vectors. (b) The electric polarization with the electric field applied along \bb\ direction.  Measurements were made with a poling field of 106\,V\,mm$^{-1}$ applied in opposite directions (red and blue curves). (c) shows the temperature dependence of the coercive field measured by neutron polarimetry.
\label{fig:charact}}
\end{figure}

Temperature profiles of magnetic and dielectric properties of our CuO crystal are presented in Fig.~\ref{fig:charact}. Figure~\ref{fig:charact}(a) shows temperature sweeps of the magnetic susceptibility recorded along the reciprocal lattice axes \ab, \bb\ and \cb. A sharp discontinuous transition is observed along all directions at $T_{\rm N1} = 213.7$\,K. Above $T_{\rm N1}$ the susceptibility increases linearly with temperature up to $T_{\rm N2} \approx 230$\,K at which point a change in slope is observed. Our data are in very good agreement with previous measurements on CuO.\cite{kimura-nature-2008} Susceptibility measurements were made on cooling and warming. We find no hysteresis at $T_{\rm N1}$ to within 0.2\,K, despite the first-order nature of this transition. We observed a similarly abrupt and non-hysteretic transition at $T_{\rm N1}$ in magnetic neutron diffraction measurements also made on cooling and warming. This is at odds with the large hysteresis of approximately 5\,K reported in diffraction measurements by Yang \textit{et al.} (Ref.~\onlinecite{yang-prb-1989}). The different behaviors may relate to slight differences in the chemical composition of the CuO crystals.

The electric polarization $P_{\rm e}$ along $\pm$\bb\ obtained by integrating the pyroelectric current with a poling field of 106\,V\,mm$^{-1}$ is shown in Fig.~\ref{fig:charact}(b). A sharp depolarization is observed close to 213\,K coincident with the AF1--AF2 magnetic transition. A maximum polarization of approximately 100\,$\mu$C\,m$^{-2}$ is recorded.

As discussed above, neutron polarimetry can be used to determine the relative proportions of the two chiral domains in CuO. Any changes in population induced by an applied electric field will affect the $\textsf{P}_{yx}$ and $\textsf{P}_{zx}$ elements in the polarization matrix while leaving the other elements unchanged.

\begin{figure}
\centering
\includegraphics[width=\columnwidth]
{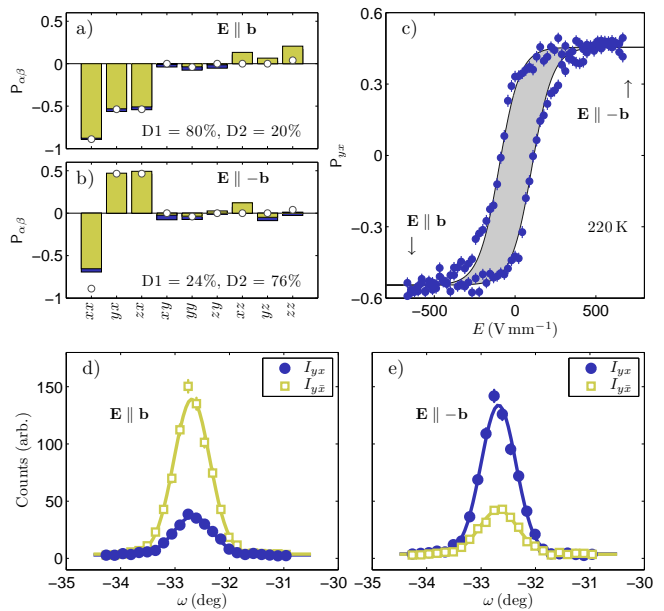}
\caption{(color online). Spherical neutron polarimetry data for CuO with an applied electric field. (a) and (b) represent the complete polarization matrix at $\qicm = (0.506,0,-0.483)$\ with applied field $E$ of $-$670\,V\,mm$^{-1}$ and $+$670\,V\,mm$^{-1}$, respectively. The color bars are the measurements, and the symbols ($\circ$) are calculated assuming the $M^{(1)}_{b}M^{(2)}_{v}$ AF2 magnetic structure and taking into account the non-ideal neutron beam polarization. The fitted populations of the two chiral magnetic domains are indicated. Panel (c) shows the $\textsf{P}_{yx}$ component of the polarization matrix during an electric field sweep at a fixed temperature of 220\,K.  Panels (d) and (e) show rocking scans through the $\qicm$ magnetic Bragg peak as measured in the $yx$ and $y\overline{x}$ polarization channels. Measurements were made at constant applied electric field of $-$670\,V\,mm$^{-1}$ (panel d) and $+$670\,V\,mm$^{-1}$ (panel e).
\label{fig:edep}}
\end{figure}

Complete polarization matrices $\textsf{P}_{\alpha\beta}$ measured with an applied external electric field of $\pm$670\,V\,mm$^{-1}$, recorded at 220\,K, are shown in Figs.~\ref{fig:edep}(a) and (b). The data clearly shows a reversal in the sign of the terms $\textsf{P}_{yx}$ and $\textsf{P}_{zx}$. A decrease in the $xx$ component in Fig.~\ref{fig:edep}(c) is anomalous. Its origin is unclear but may be due to a small misalignment of the crystal with respect to the Blume frame of reference which also causes the $\textsf{P}_{xz}$ term to be non-zero.

Measurements of the $\textsf{P}_{yx}$ component of the polarization matrix at a constant temperature of 220\,K after initial field cooling are shown in Fig.~\ref{fig:edep}(c). We find that by sweeping the electric field, which is applied in the -\bb direction, from 670\,V\,mm$^{-1}$ to $-$670\,V\,mm$^{-1}$ and back to 670\,V\,mm$^{-1}$ we observe a change in sign and magnitude of $\textsf{P}_{yx}$. It follows that the electric field must be coupled to the chiral magnetic domains. The coercive field, defined as the electric field at which the chiral domains are equally populated, is approximately 90\,V\,mm$^{-1}$ at this temperature.

We measured $\textsf{P}_{yx}$ loops as a function of electric field at a series of temperatures within the AF2 phase. The coercive field obtained from these loops, shown in Fig~\ref{fig:charact}(c), are found to increase with decreasing temperature. The system becomes softer at higher temperatures and therefore the electric field required to balance the domain population is reduced up to the point of the phase transition at 230\,K. No hysteresis and hence no coercive field is found in the AF1 or paramagnetic (PM) phases.  To verify that the electric field switching of domains exists only in the AF2 phase, hysteresis loops were also made at 240\,K (centered on $\qicm$) and 210\,K (centered on $\qicm$\ and $\qcm$). These measurements showed field-independent behavior of $\textsf{P}_{yx}$, as expected.

We note that the magnitude of $\textsf{P}_{yx}$ never reaches the maximal value of close to $\pm 1$, as expected for the case of a single chiral domain in Eq.~\ref{eq:polmat}. Even at large electric fields a significant depolarization is observed such that a saturation value of $|\textsf{P}_{yx}|\approx 0.5$, is reached at around $\pm 500$\,V\,mm$^{-1}$ when measured at 220\,K. Hysteresis loops measured at both lower and higher temperatures within the AF2 phase show similar behavior with the saturation value of $|\textsf{P}_{yx}| \approx 0.5$.

To examine this effect further we performed rocking scans through the $\qicm$\ Bragg peak. Figures~\ref{fig:edep}(d) and (e) show the intensities in the spin-flip and non-spin-flip $yx$ polarization channels at the extrema of the hysteresis loop. We would expect the left-handed helical domain (i.e.~that shown in Fig.~\ref{fig:crystal}(b)) to scatter mainly into the $yx$ polarization channel, and the right-handed helical domain to scatter mainly into the $y\bar{x}$ channel. Taking into account the non-ideal neutron beam polarization in the experiment we calculate the ratio $I_{yx}/I_{y\bar{x}}$ to be 15 for a single magnetic domain. However, from Fig.~\ref{fig:edep}(d) this ratio is closer to 3.4.

This discrepancy suggests either that the assumed magnetic structure for the AF2 phase of CuO is incorrect, or that complete switching between magnetic domains is not achieved even though $\textsf{P}_{yx}$ apparently saturates at the highest measured electric fields. The magnetic structure in the AF2 phase of CuO has been tightly constrained by previous neutron diffraction and polarimetry studies,\cite{brown-jpcm-1991, ain-jpcm-1992} as well as analysis of our own results,\cite{footnote} and so it seems reasonable to assume that the magnetic structure that we consider is indeed the correct one. Moreover, it can be shown that $\textsf{P}_{yx}^2 + \textsf{P}_{yy}^2 + \textsf{P}_{yz}^2 = 1$ for a single domain of any magnetic structure. This sum rule is not satisfied by our polarimetry data --- see Figs.~\ref{fig:edep}(a) and (b). We conclude, therefore, that electric field conversion to a single magnetic domain is never achieved in CuO.

To quantify the domain populations, we have used the model for the AF2 magnetic structure where basis vectors along $\bf b$ correspond to the $\Gamma_1$ representation and the basis vectors in ${\bf a}$-${\bf c}$ plane to $\Gamma_2$. A correction was made to the simulations to account for non-ideal beam polarization assuming a beam polarization efficiency of 94\%.  We fitted the complete measured polarization matrices allowing just the domain fraction to vary. The domains D1 and D2 correspond to left- and right-handed spin structures, respectively.  The electric field switches between the domains, with a large positive electric field yielding a predominantly left-handed domain and, conversely, a large negative field promoting the right-handed domain.

We find that typically, in the AF2 phase, at electric fields approaching saturation in $\textsf{P}_{yx}$, the magnetic domains are populated in approximately 80:20 proportion.  We have observed, therefore, that the electric field does not induce a single magnetic domain. Possible reasons for this are, (i) crystal defects acting to pin the magnetic structure and prevent full domain reversal, (ii) the high-temperature of the multiferroic phase causing thermal relaxation of the domain population in the time frame of the experiment, or (iii) a reduced electric field strength near the edges of the crystal caused by the incomplete coating of the surfaces with the gold electrodes.

\section{Discussion}

Density functional theory calculations of the electric polarization in the multiferroic phase of CuO were reported in Refs.~\onlinecite{giovannetti-prl-2011,jin-arxiv:1007.2274v3}. The magnitude of ${\bf P}_{\rm e}$ predicted by both groups is consistent with the experimental bulk value of approximately 100\,$\mu$C\,m$^{-2}$. However, we observe that the direction of ${\bf P}_{\rm e}$ is along $+{\bf b}$ for the left-handed domain, in contradiction to the direction predicted by one of the models.\cite{jin-arxiv:1007.2274v3}

The magnetically-induced ferroelectric polarization in CuO has been explained through a number of models, all of which are based upon the inverse Dzyaloshinkii--Moriya interaction between nearest-neighbor, non-collinear spins. For such spin structures, the DM interaction may reduce the system's energy, as described by the following term in the Hamiltonian,
\begin{equation}
\mathcal{H}_{\rm DM} = \sum_{i,j} \mathbf{D}_{ij} \cdot (\mathbf{S}_i\times \mathbf{S}_j).
\end{equation}
The DM vector, $\mathbf{D}_{ij}$, can be resolved into components perpendicular and parallel to the vector $\mathbf{r}_{ij}$ connecting the spins as,
\begin{equation}
\mathbf{D}_{ij} = \mathbf{P}_{ij}\times\mathbf{r}_{ij} +
\sigma_{ij} \mathbf{r}_{ij},
\label{dvec}
\end{equation}
where $\mathbf{P}_{ij}$ and $\sigma_{ij}$ are a polar vector and pseudoscalar, respectively. In CuO, $\mathbf{P}_{ij}$ is proportional to a local electric polarization, and $\sigma_{ij}$ is related to the magnetic chirality. Usually only the first term is considered when explaining the evolution of ferroelectricity, as it explicitly includes a polar vector. This is indeed the case in all previous studies of CuO. Kimura {\it et al.} (Ref.~\onlinecite{kimura-nature-2008}) consider a spin cycloid whose propagation vector is along the direction of the incommensurate modulation $\qicm-\qcm$, and Giovanetti {\it et al.} (Ref.~\onlinecite{giovannetti-prl-2011}) and Jin {\it et al.} (Ref.~\onlinecite{jin-arxiv:1007.2274v3}) find that the important magnetoelectric interaction is between approximately perpendicular spins on neighboring chains of Cu atoms running along the $[1,0,1]$ direction and does not depend on the incommensurate modulation.

The second term of Eq.~\ref{dvec} does not inherently give rise to a polarization. However, it has recently been shown that the chirality of a magnetic structure, $\sigma_{ij} $, can induce an electric polarization through coupling to a unique structural rotation.\cite{johnson-prl-2011} Furthermore, the electric polarization is then strictly constrained to lie parallel to the rotation axis. This coupling is limited to a small class of crystal structures which are termed ferroaxial.\cite{johnson-prl-2011} CuO adopts the space group $C2/c$, a member of the ferroaxial crystal class, with a two-fold rotation axis parallel to the crystallographic $b$-axis. Coupling between magnetic chirality and the crystal structure could therefore result in an electric polarization parallel to the $b$-axis, as observed. It would be of interest to include this chiral term in future DFT calculations.

A recent study of the bulk magnetoelectric properties of polycrystalline CuO showed that the application of a magnetic field has little effect on the ferroelectricity and concluded that the magnetoelectric coupling is very weak.\cite{wang-jap-2011} Our method of applying an electric field and measuring the effect on the magnetic structure shows that the magnetoelectric coupling is in fact strong. The difference between these experiments is that in our work we are probing the coupling between the macroscopic polarization and a spatially-varying magnetization, whereas in Ref.~\onlinecite{wang-jap-2011} the coupling is between the macroscopic polarization and a uniform magnetization. These studies are therefore complementary, not contradictory.

\section{Summary}
We have used polarized neutron diffraction to show that an electric field applied along the $b$ axis is able to switch between  magnetic domains in CuO. The results demonstrate that the ferroelectric polarization is directly coupled to the chiral magnetic order. We have solved the magnetic structure in the multiferroic phase by representation analysis and confirmed that it is consistent with previous data and that it supports a ferroelectric polarization. The microscopic origin of the ferroelectricity is consistent with models based on the inverse Dzyaloshinkii-Moriya mechanism.\cite{giovannetti-prl-2011,jin-arxiv:1007.2274v3, toledano-prl-2011}


\begin{acknowledgments}
We wish to acknowledge M.~Zolliker, M.~Bartkowiak, R.~Makin, Y. Bodenthin and U. Staub for important contributions relating to the technical aspects of the experiment. We are also grateful to I.~Cabrera, J.~Lorenzana, P.~G.~Radaelli and T.~Kimura for insightful discussions. PB is grateful for the provision of a studentship from the UK Engineering and Physical Sciences Research Council and Paul Scherrer Institut. This research project has been supported by the European Commission under the 7th Framework Programme through the `Research Infrastructures' action of the `Capacities' Programme, contract No: CP-CSA\_INFRA-2008-1.1.1 Number 226507-NMI3.
\end{acknowledgments}

\bibliographystyle{apsrev4-1}

\bibliography{biblio}

\begin{thebibliography}{33}%
\makeatletter
\providecommand \@ifxundefined [1]{%
 \@ifx{#1\undefined}
}%
\providecommand \@ifnum [1]{%
 \ifnum #1\expandafter \@firstoftwo
 \else \expandafter \@secondoftwo
 \fi
}%
\providecommand \@ifx [1]{%
 \ifx #1\expandafter \@firstoftwo
 \else \expandafter \@secondoftwo
 \fi
}%
\providecommand \natexlab [1]{#1}%
\providecommand \enquote  [1]{``#1''}%
\providecommand \bibnamefont  [1]{#1}%
\providecommand \bibfnamefont [1]{#1}%
\providecommand \citenamefont [1]{#1}%
\providecommand \href@noop [0]{\@secondoftwo}%
\providecommand \href [0]{\begingroup \@sanitize@url \@href}%
\providecommand \@href[1]{\@@startlink{#1}\@@href}%
\providecommand \@@href[1]{\endgroup#1\@@endlink}%
\providecommand \@sanitize@url [0]{\catcode `\\12\catcode `\$12\catcode
  `\&12\catcode `\#12\catcode `\^12\catcode `\_12\catcode `\%12\relax}%
\providecommand \@@startlink[1]{}%
\providecommand \@@endlink[0]{}%
\providecommand \url  [0]{\begingroup\@sanitize@url \@url }%
\providecommand \@url [1]{\endgroup\@href {#1}{\urlprefix }}%
\providecommand \urlprefix  [0]{URL }%
\providecommand \Eprint [0]{\href }%
\providecommand \doibase [0]{http://dx.doi.org/}%
\providecommand \selectlanguage [0]{\@gobble}%
\providecommand \bibinfo  [0]{\@secondoftwo}%
\providecommand \bibfield  [0]{\@secondoftwo}%
\providecommand \translation [1]{[#1]}%
\providecommand \BibitemOpen [0]{}%
\providecommand \bibitemStop [0]{}%
\providecommand \bibitemNoStop [0]{.\EOS\space}%
\providecommand \EOS [0]{\spacefactor3000\relax}%
\providecommand \BibitemShut  [1]{\csname bibitem#1\endcsname}%
\let\auto@bib@innerbib\@empty
\bibitem [{\citenamefont {Wang}\ \emph {et~al.}(2009)\citenamefont {Wang},
  \citenamefont {Liu},\ and\ \citenamefont {Ren}}]{wang-advphys-2009}%
  \BibitemOpen
  \bibfield  {author} {\bibinfo {author} {\bibfnamefont {K.~F.}\ \bibnamefont
  {Wang}}, \bibinfo {author} {\bibfnamefont {J.~M.}\ \bibnamefont {Liu}}, \
  and\ \bibinfo {author} {\bibfnamefont {Z.~F.}\ \bibnamefont {Ren}},\
  }\href@noop {} {\bibfield  {journal} {\bibinfo  {journal} {Adv. Phys.}\
  }\textbf {\bibinfo {volume} {58}},\ \bibinfo {pages} {321} (\bibinfo {year}
  {2009})}\BibitemShut {NoStop}%
\bibitem [{\citenamefont {Cheong}\ and\ \citenamefont
  {Mostovoy}(2007)}]{cheong-nature-2007}%
  \BibitemOpen
  \bibfield  {author} {\bibinfo {author} {\bibfnamefont {S.~W.}\ \bibnamefont
  {Cheong}}\ and\ \bibinfo {author} {\bibfnamefont {M.}~\bibnamefont
  {Mostovoy}},\ }\href@noop {} {\bibfield  {journal} {\bibinfo  {journal}
  {Nature Mater.}\ }\textbf {\bibinfo {volume} {6}},\ \bibinfo {pages} {13}
  (\bibinfo {year} {2007})}\BibitemShut {NoStop}%
\bibitem [{\citenamefont {Khomskii}(2009)}]{khomskii-physics-2009}%
  \BibitemOpen
  \bibfield  {author} {\bibinfo {author} {\bibfnamefont {D.}~\bibnamefont
  {Khomskii}},\ }\href@noop {} {\bibfield  {journal} {\bibinfo  {journal}
  {Physics}\ }\textbf {\bibinfo {volume} {2}},\ \bibinfo {pages} {20} (\bibinfo
  {year} {2009})}\BibitemShut {NoStop}%
\bibitem [{\citenamefont {Kimura}\ \emph {et~al.}(2008)\citenamefont {Kimura},
  \citenamefont {Sekio}, \citenamefont {Nakamura}, \citenamefont {Siegrist},\
  and\ \citenamefont {Ramirez}}]{kimura-nature-2008}%
  \BibitemOpen
  \bibfield  {author} {\bibinfo {author} {\bibfnamefont {T.}~\bibnamefont
  {Kimura}}, \bibinfo {author} {\bibfnamefont {Y.}~\bibnamefont {Sekio}},
  \bibinfo {author} {\bibfnamefont {H.}~\bibnamefont {Nakamura}}, \bibinfo
  {author} {\bibfnamefont {T.}~\bibnamefont {Siegrist}}, \ and\ \bibinfo
  {author} {\bibfnamefont {A.~P.}\ \bibnamefont {Ramirez}},\ }\href@noop {}
  {\bibfield  {journal} {\bibinfo  {journal} {Nature Mater.}\ }\textbf
  {\bibinfo {volume} {7}},\ \bibinfo {pages} {291} (\bibinfo {year}
  {2008})}\BibitemShut {NoStop}%
\bibitem [{\citenamefont {Brockhouse}(1954)}]{brockhouse-pr-1954}%
  \BibitemOpen
  \bibfield  {author} {\bibinfo {author} {\bibfnamefont {B.~N.}\ \bibnamefont
  {Brockhouse}},\ }\href@noop {} {\bibfield  {journal} {\bibinfo  {journal}
  {Phys. Rev.}\ }\textbf {\bibinfo {volume} {94}},\ \bibinfo {pages} {781}
  (\bibinfo {year} {1954})}\BibitemShut {NoStop}%
\bibitem [{\citenamefont {Forsyth}\ \emph {et~al.}(1988)\citenamefont
  {Forsyth}, \citenamefont {Brown},\ and\ \citenamefont
  {Wanklyn}}]{forsyth-jphysc-1988}%
  \BibitemOpen
  \bibfield  {author} {\bibinfo {author} {\bibfnamefont {J.~B.}\ \bibnamefont
  {Forsyth}}, \bibinfo {author} {\bibfnamefont {P.~J.}\ \bibnamefont {Brown}},
  \ and\ \bibinfo {author} {\bibfnamefont {B.~M.}\ \bibnamefont {Wanklyn}},\
  }\href@noop {} {\bibfield  {journal} {\bibinfo  {journal} {J. Phys. C}\
  }\textbf {\bibinfo {volume} {21}},\ \bibinfo {pages} {2917} (\bibinfo {year}
  {1988})}\BibitemShut {NoStop}%
\bibitem [{\citenamefont {A\"{i}n}\ \emph {et~al.}(1992)\citenamefont
  {A\"{i}n}, \citenamefont {Menelle}, \citenamefont {Wanklyn},\ and\
  \citenamefont {Bertaut}}]{ain-jpcm-1992}%
  \BibitemOpen
  \bibfield  {author} {\bibinfo {author} {\bibfnamefont {M.}~\bibnamefont
  {A\"{i}n}}, \bibinfo {author} {\bibfnamefont {A.}~\bibnamefont {Menelle}},
  \bibinfo {author} {\bibfnamefont {B.~M.}\ \bibnamefont {Wanklyn}}, \ and\
  \bibinfo {author} {\bibfnamefont {E.~F.}\ \bibnamefont {Bertaut}},\
  }\href@noop {} {\bibfield  {journal} {\bibinfo  {journal} {J. Phys.: Condens.
  Matter}\ }\textbf {\bibinfo {volume} {4}},\ \bibinfo {pages} {5327} (\bibinfo
  {year} {1992})}\BibitemShut {NoStop}%
\bibitem [{\citenamefont {Brown}\ \emph {et~al.}(1991)\citenamefont {Brown},
  \citenamefont {Chattopadhyay}, \citenamefont {Forsyth}, \citenamefont
  {Nunez},\ and\ \citenamefont {Tasset}}]{brown-jpcm-1991}%
  \BibitemOpen
  \bibfield  {author} {\bibinfo {author} {\bibfnamefont {P.~J.}\ \bibnamefont
  {Brown}}, \bibinfo {author} {\bibfnamefont {T.}~\bibnamefont
  {Chattopadhyay}}, \bibinfo {author} {\bibfnamefont {J.~B.}\ \bibnamefont
  {Forsyth}}, \bibinfo {author} {\bibfnamefont {V.}~\bibnamefont {Nunez}}, \
  and\ \bibinfo {author} {\bibfnamefont {F.}~\bibnamefont {Tasset}},\
  }\href@noop {} {\bibfield  {journal} {\bibinfo  {journal} {J. Phys.: Condens.
  Matter}\ }\textbf {\bibinfo {volume} {3}},\ \bibinfo {pages} {4281} (\bibinfo
  {year} {1991})}\BibitemShut {NoStop}%
\bibitem [{\citenamefont {Yang}\ \emph {et~al.}(1989)\citenamefont {Yang},
  \citenamefont {Thurston}, \citenamefont {Tranquada},\ and\ \citenamefont
  {Shirane}}]{yang-prb-1989}%
  \BibitemOpen
  \bibfield  {author} {\bibinfo {author} {\bibfnamefont {B.~X.}\ \bibnamefont
  {Yang}}, \bibinfo {author} {\bibfnamefont {T.~R.}\ \bibnamefont {Thurston}},
  \bibinfo {author} {\bibfnamefont {J.~M.}\ \bibnamefont {Tranquada}}, \ and\
  \bibinfo {author} {\bibfnamefont {G.}~\bibnamefont {Shirane}},\ }\href@noop
  {} {\bibfield  {journal} {\bibinfo  {journal} {Phys. Rev. B}\ }\textbf
  {\bibinfo {volume} {39}},\ \bibinfo {pages} {4343} (\bibinfo {year}
  {1989})}\BibitemShut {NoStop}%
\bibitem [{\citenamefont {Toledano}\ \emph {et~al.}(2011)\citenamefont
  {Toledano}, \citenamefont {Leo}, \citenamefont {Khalyavin}, \citenamefont
  {Chapon}, \citenamefont {Hoffmann}, \citenamefont {Meier},\ and\
  \citenamefont {Fiebig}}]{toledano-prl-2011}%
  \BibitemOpen
  \bibfield  {author} {\bibinfo {author} {\bibfnamefont {P.}~\bibnamefont
  {Toledano}}, \bibinfo {author} {\bibfnamefont {N.}~\bibnamefont {Leo}},
  \bibinfo {author} {\bibfnamefont {D.~D.}\ \bibnamefont {Khalyavin}}, \bibinfo
  {author} {\bibfnamefont {L.~C.}\ \bibnamefont {Chapon}}, \bibinfo {author}
  {\bibfnamefont {T.}~\bibnamefont {Hoffmann}}, \bibinfo {author}
  {\bibfnamefont {D.}~\bibnamefont {Meier}}, \ and\ \bibinfo {author}
  {\bibfnamefont {M.}~\bibnamefont {Fiebig}},\ }\href@noop {} {\bibfield
  {journal} {\bibinfo  {journal} {Phys. Rev. Lett.}\ }\textbf {\bibinfo
  {volume} {106}},\ \bibinfo {pages} {257601} (\bibinfo {year}
  {2011})}\BibitemShut {NoStop}%
\bibitem [{\citenamefont {Giovannetti}\ \emph {et~al.}(2011)\citenamefont
  {Giovannetti}, \citenamefont {Kumar}, \citenamefont {Stroppa}, \citenamefont
  {van~den Brink}, \citenamefont {Picozzi},\ and\ \citenamefont
  {Lorenzana}}]{giovannetti-prl-2011}%
  \BibitemOpen
  \bibfield  {author} {\bibinfo {author} {\bibfnamefont {G.}~\bibnamefont
  {Giovannetti}}, \bibinfo {author} {\bibfnamefont {S.}~\bibnamefont {Kumar}},
  \bibinfo {author} {\bibfnamefont {A.}~\bibnamefont {Stroppa}}, \bibinfo
  {author} {\bibfnamefont {J.}~\bibnamefont {van~den Brink}}, \bibinfo {author}
  {\bibfnamefont {S.}~\bibnamefont {Picozzi}}, \ and\ \bibinfo {author}
  {\bibfnamefont {J.}~\bibnamefont {Lorenzana}},\ }\href@noop {} {\bibfield
  {journal} {\bibinfo  {journal} {Phys. Rev. Lett.}\ }\textbf {\bibinfo
  {volume} {106}},\ \bibinfo {pages} {026401} (\bibinfo {year}
  {2011})}\BibitemShut {NoStop}%
\bibitem [{\citenamefont {Jin}\ \emph {et~al.}(2010)\citenamefont {Jin},
  \citenamefont {Cao}, \citenamefont {Guo},\ and\ \citenamefont
  {He}}]{jin-arxiv:1007.2274v3}%
  \BibitemOpen
  \bibfield  {author} {\bibinfo {author} {\bibfnamefont {G.}~\bibnamefont
  {Jin}}, \bibinfo {author} {\bibfnamefont {K.}~\bibnamefont {Cao}}, \bibinfo
  {author} {\bibfnamefont {G.-C.}\ \bibnamefont {Guo}}, \ and\ \bibinfo
  {author} {\bibfnamefont {L.}~\bibnamefont {He}},\ }\href@noop {} {\bibfield
  {journal} {\bibinfo  {journal} {arXiv:1007.2274v3}\ } (\bibinfo {year}
  {2010})}\BibitemShut {NoStop}%
\bibitem [{\citenamefont {Dzyaloshinskii}(1964)}]{dzyaloshinskii-1964}%
  \BibitemOpen
  \bibfield  {author} {\bibinfo {author} {\bibfnamefont {I.}~\bibnamefont
  {Dzyaloshinskii}},\ }\href@noop {} {\bibfield  {journal} {\bibinfo  {journal}
  {Sov. Phys. JETP}\ }\textbf {\bibinfo {volume} {19}},\ \bibinfo {pages} {960}
  (\bibinfo {year} {1964})}\BibitemShut {NoStop}%
\bibitem [{\citenamefont {Moriya}(1960)}]{moriya-pr-1960}%
  \BibitemOpen
  \bibfield  {author} {\bibinfo {author} {\bibfnamefont {T.}~\bibnamefont
  {Moriya}},\ }\href@noop {} {\bibfield  {journal} {\bibinfo  {journal} {Phys.
  Rev.}\ }\textbf {\bibinfo {volume} {120}},\ \bibinfo {pages} {91} (\bibinfo
  {year} {1960})}\BibitemShut {NoStop}%
\bibitem [{\citenamefont {Shimizu}\ \emph {et~al.}(2003)\citenamefont
  {Shimizu}, \citenamefont {Matsumoto}, \citenamefont {Goto}, \citenamefont
  {Chandrasekhar~Rao}, \citenamefont {Yoshimura},\ and\ \citenamefont
  {Kosuge}}]{shimizu-prb-2003}%
  \BibitemOpen
  \bibfield  {author} {\bibinfo {author} {\bibfnamefont {T.}~\bibnamefont
  {Shimizu}}, \bibinfo {author} {\bibfnamefont {T.}~\bibnamefont {Matsumoto}},
  \bibinfo {author} {\bibfnamefont {A.}~\bibnamefont {Goto}}, \bibinfo {author}
  {\bibfnamefont {T.~V.}\ \bibnamefont {Chandrasekhar~Rao}}, \bibinfo {author}
  {\bibfnamefont {K.}~\bibnamefont {Yoshimura}}, \ and\ \bibinfo {author}
  {\bibfnamefont {K.}~\bibnamefont {Kosuge}},\ }\href@noop {} {\bibfield
  {journal} {\bibinfo  {journal} {Phys. Rev. B}\ }\textbf {\bibinfo {volume}
  {68}},\ \bibinfo {pages} {224433} (\bibinfo {year} {2003})}\BibitemShut
  {NoStop}%
\bibitem [{\citenamefont {Ain}\ \emph {et~al.}(1989)\citenamefont {Ain},
  \citenamefont {Reichardt}, \citenamefont {Hennion}, \citenamefont {Pepy},\
  and\ \citenamefont {Wanklyn}}]{ain-physicaC-1989}%
  \BibitemOpen
  \bibfield  {author} {\bibinfo {author} {\bibfnamefont {M.}~\bibnamefont
  {Ain}}, \bibinfo {author} {\bibfnamefont {W.}~\bibnamefont {Reichardt}},
  \bibinfo {author} {\bibfnamefont {B.}~\bibnamefont {Hennion}}, \bibinfo
  {author} {\bibfnamefont {G.}~\bibnamefont {Pepy}}, \ and\ \bibinfo {author}
  {\bibfnamefont {B.~M.}\ \bibnamefont {Wanklyn}},\ }\href@noop {} {\bibfield
  {journal} {\bibinfo  {journal} {Physica C}\ }\textbf {\bibinfo {volume}
  {162}},\ \bibinfo {pages} {1279} (\bibinfo {year} {1989})}\BibitemShut
  {NoStop}%
\bibitem [{\citenamefont {Boothroyd}\ \emph {et~al.}(1997)\citenamefont
  {Boothroyd}, \citenamefont {Mukherjee}, \citenamefont {Fulton}, \citenamefont
  {Perring}, \citenamefont {Eccleston}, \citenamefont {Mook},\ and\
  \citenamefont {Wanklyn}}]{boothroyd-physicaB-1997}%
  \BibitemOpen
  \bibfield  {author} {\bibinfo {author} {\bibfnamefont {A.~T.}\ \bibnamefont
  {Boothroyd}}, \bibinfo {author} {\bibfnamefont {A.}~\bibnamefont
  {Mukherjee}}, \bibinfo {author} {\bibfnamefont {S.}~\bibnamefont {Fulton}},
  \bibinfo {author} {\bibfnamefont {T.~G.}\ \bibnamefont {Perring}}, \bibinfo
  {author} {\bibfnamefont {R.~S.}\ \bibnamefont {Eccleston}}, \bibinfo {author}
  {\bibfnamefont {H.~A.}\ \bibnamefont {Mook}}, \ and\ \bibinfo {author}
  {\bibfnamefont {B.~M.}\ \bibnamefont {Wanklyn}},\ }\href@noop {} {\bibfield
  {journal} {\bibinfo  {journal} {Physica B}\ }\textbf {\bibinfo {volume}
  {234}},\ \bibinfo {pages} {731} (\bibinfo {year} {1997})}\BibitemShut
  {NoStop}%
\bibitem [{\citenamefont {Cabrera}\ \emph {et~al.}(2009)\citenamefont
  {Cabrera}, \citenamefont {Kenzelmann}, \citenamefont {Lawes}, \citenamefont
  {Chen}, \citenamefont {Chen}, \citenamefont {Erwin}, \citenamefont {Gentile},
  \citenamefont {Leao}, \citenamefont {Lynn}, \citenamefont {Rogado},
  \citenamefont {Cava},\ and\ \citenamefont {Broholm}}]{cabrera-prl-2009}%
  \BibitemOpen
  \bibfield  {author} {\bibinfo {author} {\bibfnamefont {I.}~\bibnamefont
  {Cabrera}}, \bibinfo {author} {\bibfnamefont {M.}~\bibnamefont {Kenzelmann}},
  \bibinfo {author} {\bibfnamefont {G.}~\bibnamefont {Lawes}}, \bibinfo
  {author} {\bibfnamefont {Y.}~\bibnamefont {Chen}}, \bibinfo {author}
  {\bibfnamefont {W.~C.}\ \bibnamefont {Chen}}, \bibinfo {author}
  {\bibfnamefont {R.}~\bibnamefont {Erwin}}, \bibinfo {author} {\bibfnamefont
  {T.~R.}\ \bibnamefont {Gentile}}, \bibinfo {author} {\bibfnamefont {J.~B.}\
  \bibnamefont {Leao}}, \bibinfo {author} {\bibfnamefont {J.~W.}\ \bibnamefont
  {Lynn}}, \bibinfo {author} {\bibfnamefont {N.}~\bibnamefont {Rogado}},
  \bibinfo {author} {\bibfnamefont {R.~J.}\ \bibnamefont {Cava}}, \ and\
  \bibinfo {author} {\bibfnamefont {C.}~\bibnamefont {Broholm}},\ }\href@noop
  {} {\bibfield  {journal} {\bibinfo  {journal} {Phys. Rev. Lett.}\ }\textbf
  {\bibinfo {volume} {103}},\ \bibinfo {pages} {087201} (\bibinfo {year}
  {2009})}\BibitemShut {NoStop}%
\bibitem [{\citenamefont {Poole}\ \emph {et~al.}(2009)\citenamefont {Poole},
  \citenamefont {Brown},\ and\ \citenamefont {Wills}}]{poole-jpcs-2009}%
  \BibitemOpen
  \bibfield  {author} {\bibinfo {author} {\bibfnamefont {A.}~\bibnamefont
  {Poole}}, \bibinfo {author} {\bibfnamefont {P.~J.}\ \bibnamefont {Brown}}, \
  and\ \bibinfo {author} {\bibfnamefont {A.~S.}\ \bibnamefont {Wills}},\
  }\href@noop {} {\bibfield  {journal} {\bibinfo  {journal} {J. Phys.: Conf.
  Ser.}\ }\textbf {\bibinfo {volume} {145}},\ \bibinfo {pages} {012074}
  (\bibinfo {year} {2009})}\BibitemShut {NoStop}%
\bibitem [{\citenamefont {Seki}\ \emph {et~al.}(2008)\citenamefont {Seki},
  \citenamefont {Yamasaki}, \citenamefont {Soda}, \citenamefont {Matsuura},
  \citenamefont {Hirota},\ and\ \citenamefont {Tokura}}]{seki-prl-2008}%
  \BibitemOpen
  \bibfield  {author} {\bibinfo {author} {\bibfnamefont {S.}~\bibnamefont
  {Seki}}, \bibinfo {author} {\bibfnamefont {Y.}~\bibnamefont {Yamasaki}},
  \bibinfo {author} {\bibfnamefont {M.}~\bibnamefont {Soda}}, \bibinfo {author}
  {\bibfnamefont {M.}~\bibnamefont {Matsuura}}, \bibinfo {author}
  {\bibfnamefont {K.}~\bibnamefont {Hirota}}, \ and\ \bibinfo {author}
  {\bibfnamefont {Y.}~\bibnamefont {Tokura}},\ }\href@noop {} {\bibfield
  {journal} {\bibinfo  {journal} {Phys. Rev. Lett.}\ }\textbf {\bibinfo
  {volume} {100}},\ \bibinfo {pages} {127201} (\bibinfo {year}
  {2008})}\BibitemShut {NoStop}%
\bibitem [{\citenamefont {Yamasaki}\ \emph {et~al.}(2007)\citenamefont
  {Yamasaki}, \citenamefont {Sagayama}, \citenamefont {Goto}, \citenamefont
  {Matsuura}, \citenamefont {Hirota}, \citenamefont {Arima},\ and\
  \citenamefont {Tokura}}]{yamasaki-prl-2007}%
  \BibitemOpen
  \bibfield  {author} {\bibinfo {author} {\bibfnamefont {Y.}~\bibnamefont
  {Yamasaki}}, \bibinfo {author} {\bibfnamefont {H.}~\bibnamefont {Sagayama}},
  \bibinfo {author} {\bibfnamefont {T.}~\bibnamefont {Goto}}, \bibinfo {author}
  {\bibfnamefont {M.}~\bibnamefont {Matsuura}}, \bibinfo {author}
  {\bibfnamefont {K.}~\bibnamefont {Hirota}}, \bibinfo {author} {\bibfnamefont
  {T.}~\bibnamefont {Arima}}, \ and\ \bibinfo {author} {\bibfnamefont
  {Y.}~\bibnamefont {Tokura}},\ }\href@noop {} {\bibfield  {journal} {\bibinfo
  {journal} {Phys. Rev. Lett.}\ }\textbf {\bibinfo {volume} {98}},\ \bibinfo
  {pages} {147204} (\bibinfo {year} {2007})}\BibitemShut {NoStop}%
\bibitem [{\citenamefont {Fabrizi}\ \emph {et~al.}(2009)\citenamefont
  {Fabrizi}, \citenamefont {Walker}, \citenamefont {Paolasini}, \citenamefont
  {de~Bergevin}, \citenamefont {Boothroyd}, \citenamefont {Prabhakaran},\ and\
  \citenamefont {McMorrow}}]{fabrizi-prl-2009}%
  \BibitemOpen
  \bibfield  {author} {\bibinfo {author} {\bibfnamefont {F.}~\bibnamefont
  {Fabrizi}}, \bibinfo {author} {\bibfnamefont {H.~C.}\ \bibnamefont {Walker}},
  \bibinfo {author} {\bibfnamefont {L.}~\bibnamefont {Paolasini}}, \bibinfo
  {author} {\bibfnamefont {F.}~\bibnamefont {de~Bergevin}}, \bibinfo {author}
  {\bibfnamefont {A.~T.}\ \bibnamefont {Boothroyd}}, \bibinfo {author}
  {\bibfnamefont {D.}~\bibnamefont {Prabhakaran}}, \ and\ \bibinfo {author}
  {\bibfnamefont {D.~F.}\ \bibnamefont {McMorrow}},\ }\href@noop {} {\bibfield
  {journal} {\bibinfo  {journal} {Phys. Rev. Lett.}\ }\textbf {\bibinfo
  {volume} {102}},\ \bibinfo {pages} {237205} (\bibinfo {year}
  {2009})}\BibitemShut {NoStop}%
\bibitem [{\citenamefont {Radaelli}\ \emph {et~al.}(2008)\citenamefont
  {Radaelli}, \citenamefont {Chapon}, \citenamefont {Daoud-Aladine},
  \citenamefont {Vecchini}, \citenamefont {Brown}, \citenamefont {Chatterji},
  \citenamefont {Park},\ and\ \citenamefont {Cheong}}]{radaelli-prl-2008}%
  \BibitemOpen
  \bibfield  {author} {\bibinfo {author} {\bibfnamefont {P.~G.}\ \bibnamefont
  {Radaelli}}, \bibinfo {author} {\bibfnamefont {L.~C.}\ \bibnamefont
  {Chapon}}, \bibinfo {author} {\bibfnamefont {A.}~\bibnamefont
  {Daoud-Aladine}}, \bibinfo {author} {\bibfnamefont {C.}~\bibnamefont
  {Vecchini}}, \bibinfo {author} {\bibfnamefont {P.~J.}\ \bibnamefont {Brown}},
  \bibinfo {author} {\bibfnamefont {T.}~\bibnamefont {Chatterji}}, \bibinfo
  {author} {\bibfnamefont {S.}~\bibnamefont {Park}}, \ and\ \bibinfo {author}
  {\bibfnamefont {S.~W.}\ \bibnamefont {Cheong}},\ }\href@noop {} {\bibfield
  {journal} {\bibinfo  {journal} {Phys. Rev. Lett.}\ }\textbf {\bibinfo
  {volume} {101}},\ \bibinfo {pages} {067205} (\bibinfo {year}
  {2008})}\BibitemShut {NoStop}%
\bibitem [{\citenamefont {Prabhakaran}\ and\ \citenamefont
  {Boothroyd}(2003)}]{prabhak-jcg-2003}%
  \BibitemOpen
  \bibfield  {author} {\bibinfo {author} {\bibfnamefont {D.}~\bibnamefont
  {Prabhakaran}}\ and\ \bibinfo {author} {\bibfnamefont {A.~T.}\ \bibnamefont
  {Boothroyd}},\ }\href@noop {} {\bibfield  {journal} {\bibinfo  {journal} {J.
  Crys. Growth}\ }\textbf {\bibinfo {volume} {250}},\ \bibinfo {pages} {77}
  (\bibinfo {year} {2003})}\BibitemShut {NoStop}%
\bibitem [{\citenamefont {Semadeni}\ \emph {et~al.}(2001)\citenamefont
  {Semadeni}, \citenamefont {Roessli},\ and\ \citenamefont
  {B\"{o}ni}}]{semadeni-physicab-2001}%
  \BibitemOpen
  \bibfield  {author} {\bibinfo {author} {\bibfnamefont {F.}~\bibnamefont
  {Semadeni}}, \bibinfo {author} {\bibfnamefont {B.}~\bibnamefont {Roessli}}, \
  and\ \bibinfo {author} {\bibfnamefont {P.}~\bibnamefont {B\"{o}ni}},\
  }\href@noop {} {\bibfield  {journal} {\bibinfo  {journal} {Physica B}\
  }\textbf {\bibinfo {volume} {297}},\ \bibinfo {pages} {152} (\bibinfo {year}
  {2001})}\BibitemShut {NoStop}%
\bibitem [{\citenamefont {Fischer}(1997)}]{fischer-physicab-1997}%
  \BibitemOpen
  \bibfield  {author} {\bibinfo {author} {\bibfnamefont {W.~E.}\ \bibnamefont
  {Fischer}},\ }\href@noop {} {\bibfield  {journal} {\bibinfo  {journal}
  {Physica B}\ }\textbf {\bibinfo {volume} {234}},\ \bibinfo {pages} {1202}
  (\bibinfo {year} {1997})}\BibitemShut {NoStop}%
\bibitem [{\citenamefont {Yamada}\ \emph {et~al.}(2004)\citenamefont {Yamada},
  \citenamefont {Zheng}, \citenamefont {Soejima},\ and\ \citenamefont
  {Kawaminami}}]{yamada-prb-2004}%
  \BibitemOpen
  \bibfield  {author} {\bibinfo {author} {\bibfnamefont {H.}~\bibnamefont
  {Yamada}}, \bibinfo {author} {\bibfnamefont {X.~G.}\ \bibnamefont {Zheng}},
  \bibinfo {author} {\bibfnamefont {Y.}~\bibnamefont {Soejima}}, \ and\
  \bibinfo {author} {\bibfnamefont {M.}~\bibnamefont {Kawaminami}},\
  }\href@noop {} {\bibfield  {journal} {\bibinfo  {journal} {Phys. Rev. B}\
  }\textbf {\bibinfo {volume} {69}},\ \bibinfo {pages} {104104} (\bibinfo
  {year} {2004})}\BibitemShut {NoStop}%
\bibitem [{\citenamefont {Bertaut}(1971)}]{bertaut-jpc-1971}%
  \BibitemOpen
  \bibfield  {author} {\bibinfo {author} {\bibfnamefont {E.~F.}\ \bibnamefont
  {Bertaut}},\ }\href@noop {} {\bibfield  {journal} {\bibinfo  {journal} {J.
  Phys. Colloques}\ }\textbf {\bibinfo {volume} {32C1}},\ \bibinfo {pages}
  {462} (\bibinfo {year} {1971})}\BibitemShut {NoStop}%
\bibitem [{foo()}]{footnote}%
  \BibitemOpen
  \href@noop {} {}\bibinfo {note} {We have tested these structures against the
  data of Brown {\it et al.}\cite{brown-jpcm-1991} Assuming a circular envelope
  for the spins and an equal population of the two chiral domains we obtain the
  following fits: $M^{(1)}_{b}M^{(1)}_{v}$, $\chi^2 \approx 180 $;
  $M^{(2)}_{b}M^{(2)}_{v}$, $\chi^2 \approx 870$. For comparison, we find for
  $M^{(1)}_{b}M^{(2)}_{v}$, $\chi^2 \approx 5.4$.}\BibitemShut {Stop}%
\bibitem [{\citenamefont {Blume}(1963)}]{blume-pr-1963}%
  \BibitemOpen
  \bibfield  {author} {\bibinfo {author} {\bibfnamefont {M.}~\bibnamefont
  {Blume}},\ }\href@noop {} {\bibfield  {journal} {\bibinfo  {journal} {Phys.
  Rev.}\ }\textbf {\bibinfo {volume} {130}},\ \bibinfo {pages} {1670} (\bibinfo
  {year} {1963})}\BibitemShut {NoStop}%
\bibitem [{\citenamefont {Izyumov}\ and\ \citenamefont
  {Maleyev}(1962)}]{izyumov-sovphys-1962}%
  \BibitemOpen
  \bibfield  {author} {\bibinfo {author} {\bibfnamefont {Y.}~\bibnamefont
  {Izyumov}}\ and\ \bibinfo {author} {\bibfnamefont {S.}~\bibnamefont
  {Maleyev}},\ }\href@noop {} {\bibfield  {journal} {\bibinfo  {journal} {Sov.
  Phys. JETP}\ }\textbf {\bibinfo {volume} {14}},\ \bibinfo {pages} {1668}
  (\bibinfo {year} {1962})}\BibitemShut {NoStop}%
\bibitem [{\citenamefont {Johnson}\ \emph {et~al.}(2011)\citenamefont
  {Johnson}, \citenamefont {Nair}, \citenamefont {Chapon}, \citenamefont
  {Bombardi}, \citenamefont {Vecchini}, \citenamefont {Prabhakaran},
  \citenamefont {Boothroyd},\ and\ \citenamefont
  {Radaelli}}]{johnson-prl-2011}%
  \BibitemOpen
  \bibfield  {author} {\bibinfo {author} {\bibfnamefont {R.~D.}\ \bibnamefont
  {Johnson}}, \bibinfo {author} {\bibfnamefont {S.}~\bibnamefont {Nair}},
  \bibinfo {author} {\bibfnamefont {L.~C.}\ \bibnamefont {Chapon}}, \bibinfo
  {author} {\bibfnamefont {A.}~\bibnamefont {Bombardi}}, \bibinfo {author}
  {\bibfnamefont {C.}~\bibnamefont {Vecchini}}, \bibinfo {author}
  {\bibfnamefont {D.}~\bibnamefont {Prabhakaran}}, \bibinfo {author}
  {\bibfnamefont {A.~T.}\ \bibnamefont {Boothroyd}}, \ and\ \bibinfo {author}
  {\bibfnamefont {P.~G.}\ \bibnamefont {Radaelli}},\ }\href@noop {} {\bibfield
  {journal} {\bibinfo  {journal} {Phys. Rev. Lett.}\ }\textbf {\bibinfo
  {volume} {107}},\ \bibinfo {pages} {137205} (\bibinfo {year}
  {2011})}\BibitemShut {NoStop}%
\bibitem [{\citenamefont {Wang}\ \emph {et~al.}(2011)\citenamefont {Wang},
  \citenamefont {Zou}, \citenamefont {Liu}, \citenamefont {Yan},\ and\
  \citenamefont {Sun}}]{wang-jap-2011}%
  \BibitemOpen
  \bibfield  {author} {\bibinfo {author} {\bibfnamefont {F.}~\bibnamefont
  {Wang}}, \bibinfo {author} {\bibfnamefont {T.}~\bibnamefont {Zou}}, \bibinfo
  {author} {\bibfnamefont {Y.}~\bibnamefont {Liu}}, \bibinfo {author}
  {\bibfnamefont {L.~Q.}\ \bibnamefont {Yan}}, \ and\ \bibinfo {author}
  {\bibfnamefont {Y.}~\bibnamefont {Sun}},\ }\href@noop {} {\bibfield
  {journal} {\bibinfo  {journal} {J. Appl. Phys.}\ }\textbf {\bibinfo {volume}
  {110}},\ \bibinfo {pages} {054106} (\bibinfo {year} {2011})}\BibitemShut
  {NoStop}%
\end{thebibliography}%

\end{document}